\documentclass[prl,twocolumn,showpacs,superscriptaddress,floatfix]{revtex4}

\usepackage{amsmath}
\usepackage{graphicx}

\begin{document}

\title{Effects of metallic contacts on electron transport through graphene}

\author{Salvador Barraza-Lopez}
\affiliation{School of Physics, Georgia Institute of Technology,
Atlanta, Georgia 30332, USA}

\author{Mihajlo Vanevi{\'c}}
\affiliation{School of Physics, Georgia Institute of Technology,
Atlanta, Georgia 30332, USA}
\affiliation{Kavli Institute of Nanoscience, Delft University of Technology,
2628 CJ Delft, The Netherlands}

\author{Markus Kindermann}
\affiliation{School of Physics, Georgia Institute of Technology,
Atlanta, Georgia 30332, USA}

\author{M. Y. Chou}
\affiliation{School of Physics, Georgia Institute of Technology,
Atlanta, Georgia 30332, USA}

\date{\today}

\begin{abstract}
We report on a first-principles study of the conductance through graphene suspended
between Al contacts as a function of junction length, width, and orientation. The
charge transfer at the leads and into the freestanding section gives rise to an
electron-hole asymmetry in the conductance and in sufficiently long junctions
induces two conductance minima at the energies of the Dirac points for suspended
and clamped regions, respectively. We obtain the potential profile along a junction
caused by doping and provide parameters for effective model calculations of the
junction conductance with weakly interacting metallic leads.
\end{abstract}


\maketitle

\noindent\emph{Introduction.} Graphene is a two-dimensional allotrope
of carbon with atoms arranged in a honeycomb lattice.
Successful fabrication of graphene monolayers by means of
mechanical exfoliation of graphite \cite{Novoselov:04}
or epitaxial growth on silicon carbide \cite{Berger:04}
has ignited tremendous interest in this material \cite{CastroNetoRMP:09}.
 Its favorable characteristics
such as chemical inertness, low dimensionality, extremely high
mobility, and easy control of carriers by applied gate voltages,
along with patterning using nanolithography,
open possibilities for further miniaturization  of devices and
the emergence of a carbon-based ``post-silicon''
electronics \cite{Berger:04,Avouris:07}.

 The electron transport at the nanometer scale is significantly affected by the contacts. The role of metallic
leads in determining the transport properties of graphene-based junctions has been addressed by
several theoretical \cite{TworzydloBeenakkerPRL96-06,%
RobinsonSchomerusPRB76-07,BlanterMartinPRB76-07,%
MojaradDattaPRB79-09,NemecTomanekCunibertiPRB77-08,
RanGuanWangYuAPL94-09}
and experimental \cite{HeerscheVandersypenMorpurgoNATURE446-07,%
LeeBurghardKernNATURENANOTECHNOLOGY3-08,%
HuardStanderGoldhaberGordonPRB78-08,%
DanneauMorpurgoPRL100-08,Geim09}
studies. Yet, a parameter-free description of transport in these systems is still lacking.
Previously, the leads were modeled as infinitely doped
graphene regions that support a large number of
propagating modes \cite{TworzydloBeenakkerPRL96-06}.
In this model, the important parameter is the ratio $W/L$
between the width ($W$) and length ($L$) of a junction:
at the Dirac point, the universal, diffusive-like regime is reached for $W/L\gg 1$.
Additional studies within a tight-binding approach have been reported, in which
either the leads form a square
lattice \cite{RobinsonSchomerusPRB76-07,%
BlanterMartinPRB76-07},
or the coupling to the leads is modeled via energy-level
broadening \cite{MojaradDattaPRB79-09}.

In this Letter, we report on large-scale first-principles calculations of electron
transport in suspended graphene taking into account the effects of metallic leads.
We study a non-magnetic junction made of graphene
contacted underneath by two aluminum (Al) leads with a small in-plane mismatch of less than 1\%.
This junction is a prototype system with weak interaction
between the metal contact and graphene, with no covalent bonds formed. Our
work represents the first quantitative study of electron transport through metal-graphene
 junctions to examine in detail previous models \cite{TworzydloBeenakkerPRL96-06,
MojaradDattaPRB79-09} with dimensions relevant to
 experiment \cite{HeerscheVandersypenMorpurgoNATURE446-07,%
LeeBurghardKernNATURENANOTECHNOLOGY3-08,%
HuardStanderGoldhaberGordonPRB78-08,%
DanneauMorpurgoPRL100-08},
up to 100 nm wide and 13.6 nm long with various $W/L$ ratios.

 The difference in the work functions of the metal and graphene
leads to the charge transfer and doping of the graphene
layer \cite{GiovannettiKhomyakovKellyPRL101-08}.
We obtain the corresponding potential profile generated by
the doping of the graphene layer along the junction. Not surprisingly, the potential
 in the region contacted by the metal starts deviating from its bulk value
before reaching the geometrical edge.
Finite doping results in specific transport features for long junctions:
two conductance minima appear at the energies of the Dirac points
of graphene in the clamped and suspended regions, respectively.
For shorter junctions, where the two minima cannot be
 resolved, an electron-hole asymmetry in the conductance is still appreciable.
The impact on transport of the orientation of graphene in the junctions is found to be negligible away
from the Dirac point. Using the potential profile obtained from our first-principles calculations and a small energy
broadening in the self-energies at the leads, we demonstrate that
a $\pi$-electron tight-binding model can accurately reproduce our first-principles transport results.
This enables us to predict the conductance for graphene junctions
with leads made of other metals, and we present results for the conductance of junctions with Au leads.

We perform transport calculations at zero source-drain bias
using the nonequilibrium Green's function (NEGF)
SMEAGOL code \cite{RochaLambertSanvitoNATUREMAT4-05}, which is interfaced with the
density-functional-theory (DFT) SIESTA
package \cite{SolerArtachoSanchezPortalJPHYSCONDENSMATTER14-02}.
 We employ norm-conserving pseudopotentials \cite{TroullierMartinsPRB43-91}
in the local-density approximation (LDA) \cite{PerdewZungerPRB23-81},
and a real-space grid equivalent to an energy cutoff of 310 Ry.
We explicitly construct fine-tuned basis sets
for C and Al atoms following the prescription of Junquera et al. \cite{JunqueraPazArtachoPRB64-01}.
The largest number of atoms (numerical orbitals) included
 in the present calculations is 464 (5600).
We made modifications to the transport code to improve memory allocation
and parallelization, in order to handle calculations of this scale.

\begin{figure}[tb]
\includegraphics[width=0.45\textwidth]{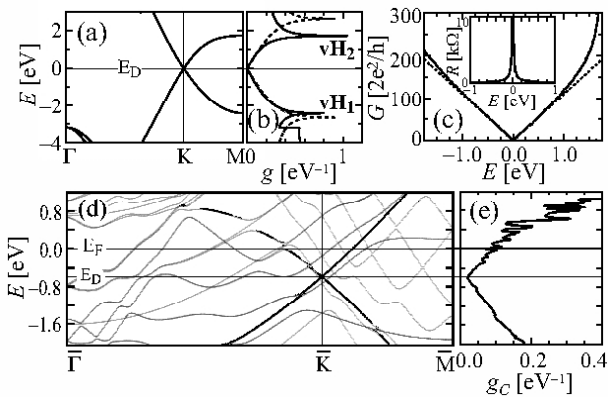}
 \caption{(a) Band structure and (b) density of states (DOS) for graphene
near the Dirac point ($E_D=0$). The two van Hove singularities are marked.
(c) Conductance of a graphene ribbon $98.1$ nm wide without metallic leads.
Inset: The resistance $R$ diverges at the Dirac point.
Dashed lines in (b) and (c) are results of tight-binding
calculations. (d) Band structure and (e) projected DOS on carbon atoms for graphene on Al. The highlighted lines in (d) are graphene bands.
}\label{fig:fig1}
\end{figure}

\noindent\emph{Graphene junctions without metal contacts.} First we discuss 
pristine graphene with no metallic leads attached. The band structure and density of states (DOS)
in the vicinity of the Dirac point ($E_D$) are shown
in Figs.~\ref{fig:fig1}(a) and (b), respectively.
The DOS around $E_D$ can be fitted to $g(E)=D|E-E_D|$ with $D=0.11$/(eV$^2$ unit cell).
In all transport calculations described in this paper, an effective ribbon width is determined by
$W=n_kw_0$, where $n_k$ is the number of $\mathbf{k}$-points
used in the calculation and $w_0$ is the size of the unit cell in the transverse ($x$) direction.
We show in Fig.~\ref{fig:fig1}(c) the conductance $G$ as a function of electron energy for a graphene ribbon of $W \approx 100$ nm.
At this width, the quantized plateaus in the conductance become unresolvable within
the energy resolution employed (0.01 eV), and $G$ resembles the linear behavior
of the DOS around $E_D$. The boundary
conditions and type of edge -- zigzag or armchair -- have an impact on the conductance only
at the vicinity of the Dirac point, in
a narrow energy range inversely proportional to the ribbon width. Away from
$E_D$, the conductance remains unaffected by this choice 
with $W$ being the same as the geometric width in the widest
ribbons studied.
 Tight-binding results with a nearest-neighbor
hopping parameter $\gamma=-$2.65 eV determined from the LDA bandstructure
are shown by dashed lines in Figs.~\ref{fig:fig1}(b) and (c) for comparison.
The agreement with the DFT results is quite good within $\pm 1$ eV of the Dirac point.
\begin{figure}[tb]
\includegraphics[width=0.4\textwidth]{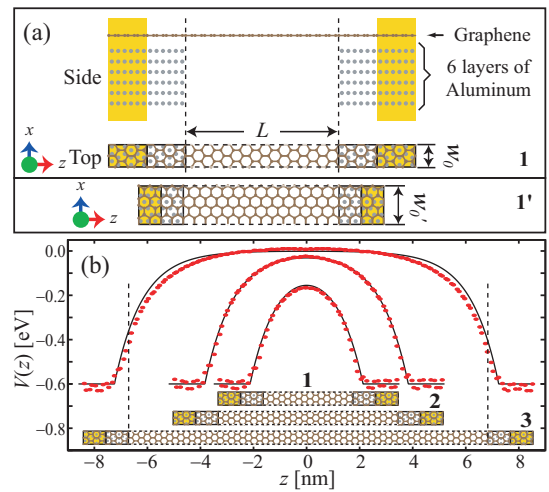}
 \caption{(Color online) (a) Schematics of Al-graphene junctions. The semi-infinite leads, shown in the shaded regions, include both graphene and Al.
The edges of the ribbon are of either armchair ({\bf 1}) or zigzag ({\bf 1'}). The unit cell sizes along the $x$-direction are $w_0=0.49$ nm and $w_0'=0.85$ nm, respectively.
$L$ is the distance between Al atoms at opposite sides of the junction. (b) Energy location of the Dirac point with respect to the Fermi level across the junctions. Junctions {\bf 1}, {\bf 2}, and {\bf 3}
have a fixed width $W=98.1$ nm, and lengths $L=3.40$ nm, $6.80$ nm, and $13.60$ nm, respectively. The DFT results (dots) are shown along with fitted curves (see text).
}\label{fig:fig2}
\end{figure}

\begin{figure}
\includegraphics[width=0.4\textwidth]{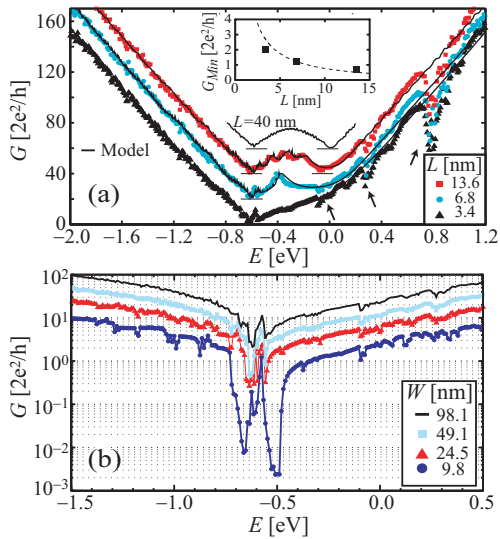}
\caption{(Color online)
(a) Conductance of junctions with a width of $W=98.1$ nm and different lengths $L$. The
curves are offset by $20\times 2e^2/h$ for clarity.
Two prominent conductance minima at $E_D$ ($-0.6$ eV) and $E_F$ (set to zero) develop as $L$
increases. The symbols represent first-principles results, and the solid lines are from model calculations (see text). Inset: minimum conductance versus $L$ (the dashed line is a guide to the eye). 
(b) Conductance of junctions with a length of $L=3.4$ nm and different widths $W$, highlighting the opening of a conduction gap as $W$ decreases and a sharp peak pinned at $E_D=-0.6$ eV.
}\label{fig:fig3}
\end{figure}

\noindent\emph{Junctions with Al-graphene contacts.} The electronic structure of graphene in
contact with Al (henceforth referred to as the lead) shows a certain
amount of hybridization arising from the interaction between graphene and the metal [Fig.~\ref{fig:fig1}(d)].
The leads are modeled by six layers of Al in the (111) orientation with
a graphene monolayer placed at a distance of 0.34
nm away, in the minimal-energy configuration \cite{GiovannettiKhomyakovKellyPRL101-08}.
Given the relative position of the Dirac point $E_D$
with respect to the Fermi level $E_F$ at the lead, $\Delta=E_D-E_F \simeq -0.6$ eV,
graphene becomes $n$-doped in the contact region.
The DOS projected on C atoms at the leads $g_C$ is nonzero at
the Dirac point with a finite value of $g_C(E_D)\sim
2\times 10^{-2}$ electrons/eV per graphene unit cell [Fig.~\ref{fig:fig1}(e)]. The fluctuations
of $g_C$ above $E_F$ are due to hybridization.

The junctions studied are schematically shown in Fig.~\ref{fig:fig2}. The leads are displayed within the shaded areas.
The Hamiltonian, overlap, and density matrices of the
leads are obtained self-consistently from DFT bulk-calculations
 and used to set up the electron density, chemical potential, and self-energies for
the transport calculations. Since the leads include graphene,
the contact area between metal and graphene is essentially infinite.
This enhances the transparency \cite{NemecTomanekCunibertiPRB77-08}, eliminating the large contact
resistance
 that would arise if charge carriers were to tunnel from Al to graphene.
An additional number of Al atoms between the leads and the freestanding section are added
so that the self-consistent electrostatic potential across the junction develops
smoothly.
 We have investigated various configurations to understand the effects
of the length ($L$), width ($W$), and orientation of the graphene ribbon on electron transport.
Junctions {\bf 1}, {\bf 1'}, {\bf 2} and {\bf 3}
have lengths  $L=3.40$ nm, $3.44$ nm, $6.80$ nm, and $13.60$ nm, respectively.
The effect of varying $L$ is studied
through junctions
{\bf 1}, {\bf 2}, and {\bf 3} with a fixed width
of $W=98.1$ nm [see Fig.~\ref{fig:fig2}(b) for geometries];
 the conductance $G$ as a function of $W$ is studied for
 junction {\bf 1} ($L=3.40$ nm); and the effect of ribbon orientation
  (armchair versus zigzag edges) is
studied for junctions {\bf 1} and {\bf 1'} by choosing
a similar width for them: $W=98.1$ nm ({\bf 1}) and $W'=98.6$ nm ({\bf 1'}).
(The potential profiles in Fig.~\ref{fig:fig2}(b) will be discussed later.)


 The conductance for junctions {\bf 1}--{\bf 3}
(with a fixed width of $W=98.1$ nm) is shown in Fig.~\ref{fig:fig3}(a).
One notices the lack of symmetry with respect to both the Fermi level (energy zero) and $E_D$ in the lead ($-0.6$ eV)
in all cases regardless of the particular values of $L$, in contrast to the symmetric curve for pristine graphene in Fig.~\ref{fig:fig1}(c)
and those displayed in Ref.~\cite{TworzydloBeenakkerPRL96-06}. Similar
asymmetries in the conductance have been seen experimentally, e.g. Refs.~\cite{HuardStanderGoldhaberGordonPRB78-08}
and \cite{FarmerPerebeinosAvourisNANOLETTERS9-09}.
In addition to the conductance minimum at $E_D$, the emergence of a second one at $E_F$ is apparent, and it becomes more prominent as the length increases.
This is because, as $L\to \infty$, the electron DOS approaches $g(E)$ at the freestanding part of the
junction and the conductance becomes that of two resistors in series \cite{FarmerPerebeinosAvourisNANOLETTERS9-09}:
$G(E)\propto g(E)g_C(E)/[g(E)+g_C(E)]$, where $g_C(E)\propto D|E-\Delta |$ is the projected DOS of the clamped region.
 Based on the same underlying physics, a double peak in resistance was experimentally demonstrated for graphene
devices comprising two regions with noticeably different doping levels \cite{FarmerLinAvourisAPL94-09}.
 The two conductance minima were also observed in junctions with gate-tunable barriers \cite{Others}.
The inset in Fig.~\ref{fig:fig3}(a) shows the minimal conductance near $E_D$ for the values of $L$
studied; it is of the order of $2e^2/h$.
A small peak at $E_D$ ($-0.6$ eV) is visible for the two shorter junctions {\bf 1} and {\bf 2}, and
becomes unnoticeable for the longer junction {\bf 3}. This small peak
is due to states that exist only in the contact areas through
hybridization with Al atoms and decreases rapidly with increasing junction length.

\begin{figure}
\includegraphics[width=0.45\textwidth]{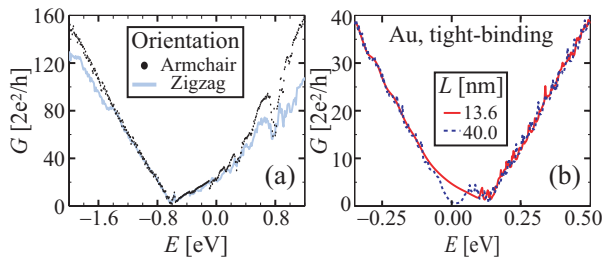}
\caption{(Color online) (a) Conductance for junctions {\bf 1} (armchair edges, $L=3.40$ nm, $W=98.1$ nm, light line) and {\bf 1'} (zigzag edges, $L=3.44$ nm, $W=98.6$ nm, dark dots).
(b) Conductance for junctions consisting of graphene with Au leads for two values of $L$ from model calculations (see text). The junction width is $W=98.1$ nm.}\label{fig:fig4}
\end{figure}

In Fig.~\ref{fig:fig3}(b) the conductance is plotted for the
shortest junction ($L=3.4$ nm) with four values of width $W$: 9.8 nm, 24.5 nm, 49.1 nm, and 98.1 nm.
The logarithmic scale of the conductance
helps examine the linear dependence of the conductance on $W$ for energies sufficiently away from $E_D$.
As mentioned above, the peak at $-0.6$ eV only appears in very short junctions. This is yet to be
observed in experiment, where longer ribbons are
customarily employed \cite{HanOzyilmazKimPRL98-07}. Set aside this peak, the
opening of a conduction gap with decreasing $W$ is evident, as well as the significant and nonlinear
drop in conductance for the narrowest junction with $W=9.8$ nm,
in full accordance with experiment \cite{HanOzyilmazKimPRL98-07}.
To investigate the effect of the edge orientation, we study the conductance of
armchair ({\bf 1}) and zigzag ({\bf 1'}) junctions of comparable dimensions [Fig.~\ref{fig:fig4}(a)].
Only small deviations are found resulting from the anisotropy of the band structure.
The effect of orientation is quite small in the vicinity of $E_D$.

\noindent\emph{Tight-binding model.} We next present a tight-binding model that can be used
to study junctions bigger in size and with arbitrary nonbonding metal contacts.
To that end, we estimate the effective potential profile $V(z)$ at the atomic scale for $\pi$-electrons along the junction, directly related
to the spatial dependence of doping created by the metallic leads. 
This profile can be obtained semiclassically from DFT as the value of the energy at the Dirac point with respect to the Fermi level, as a function of $z$: $V(z)=E_D(z) - E_F$. We extract this quantity from the position-dependent energy shift of
the van Hove singularities closest to the Dirac point
[denoted by vH$_1$ and vH$_2$ in Fig.~\ref{fig:fig1}(b)] along the junction.
The results for junctions {\bf 1}--{\bf 3} are shown in Fig.~\ref{fig:fig2}(b).
Significant doping occurs at distances up to only a few nanometers from the edge of metallic leads.
For the two longer junctions, {\bf 2} ($L=$6.80 nm) and {\bf 3} ($L=$13.60 nm),
$E_D$ is very close to $E_F$ in the middle of the freestanding section ($z \sim 0$).
  We fitted our DFT results for $V(z)$ by the following expression [solid lines in Fig.~\ref{fig:fig2}(b)]:
\begin{equation}
V(z) =
\begin{cases}
\Delta\,  \cosh(z/\lambda)/ \cosh(L_{\rm eff}/2\lambda)&
\text{for }|z|<{L_{\rm eff}}/{2}\,,\\
\Delta  &\text{otherwise}.
 \end{cases}
\end{equation}
The parameters used are $\Delta=-0.6$ eV, $\lambda=1.05$ nm and $L_{\rm eff}=L+5a_0$, where $a_0=0.142$ nm is the
C-C interatomic distance. We emphasize that on the sub-nanometer scale,
$V(z)$ starts to deviate from its asymptotic value inside the region of intermediate Al atoms,
before reaching the atomic boundary of the contact indicated by the
vertical dashed line in Fig.~\ref{fig:fig2}(b). In our calculations, sufficiently many Al 
atoms are included
between the bulk (shaded) leads and the freestanding section in order to map the transitional behavior
of $V(z)$.
This profile provides a refinement of models where
sharp steps at the metal edge are assumed \cite{TworzydloBeenakkerPRL96-06,LeeBurghardKernNATURENANOTECHNOLOGY3-08} and
will become relevant as experimental junctions shrink in size.

We have performed conductance calculations within a tight-binding
approach using the potential profile $V(z)$
and a smearing $\delta=8$ meV introduced through the leads self-energies \cite{MojaradDattaPRB79-09} (to
account for the nonzero conductance at $E_D$ and $E_F$ and the small peak at $E_D$ in Fig.~\ref{fig:fig3}).
We obtain excellent agreement with our full-scale calculations [as indicated by the
solid lines in Fig.~\ref{fig:fig3}(a)], with the exception of sharp dips
in the conductance [indicated by arrows in Fig.~\ref{fig:fig3}(a)] which
are not reproduced. These dips occur at the anticrossings energies due to
hybridization with aluminum [cf.~band structure in Fig.~\ref{fig:fig1}(d)].
We have also used the model to calculate the conductance for a
longer junction ($L=40$ nm), where the two conductance minimum are fully developed.
In the limit $\Delta \to \infty$ only one conductance minimum appears
and as a consequence the electron-hole symmetry is preserved \cite{TworzydloBeenakkerPRL96-06}.

\noindent\emph{Junctions with other nonbonding metal contacts.} We use the model presented above to make a prediction for the conductance with
Au as the metal lead. In this case $\Delta =+0.12$ eV and graphene is $p$-doped \cite{GiovannettiKhomyakovKellyPRL101-08}.
Using the same values of $\lambda$, $L_{\rm eff}$ in Eq.~(1) and the smearing parameter $\delta$ as before, the conductance from tight-binding calculations
is shown in Fig.~\ref{fig:fig4}(b) for two junctions with a fixed width of $W=98.1$ nm.
The solid line shows results for $L=13.6$ nm and the dashed line for $L=40$ nm.
 For small values of $\Delta$, the two
conductance minimum may be smeared out if the amount of charge
fluctuations is significant \cite{LeeBurghardKernNATURENANOTECHNOLOGY3-08, Geim09,
RossiAdamDasSarmaPRB79-09}. Nevertheless, the
electron-hole asymmetry should remain noticeable in the conductance curve.
 Our results give microscopic justification to prior theoretical studies where graphene junctions
 are modeled as an isolated graphene sheet with a space-dependent
 potential \cite{TworzydloBeenakkerPRL96-06}.
  However, we have demonstrated that to obtain accurate results, one needs to take into
  account a small energy broadening \cite{MojaradDattaPRB79-09} and allow for a finite potential
  at the contacts.

\noindent\emph{Conclusion.} We have performed transport calculations for graphene
junctions attached to Al leads using first-principles nonequilibrium Green's function
methods. The conductance features vary with the length and width of the junction, but
are less sensitive to the ribbon orientation. We show that nonbonding
metallic leads induce a noticeable
electron-hole asymmetry in the conductance. The opening of a conduction gap with decreasing $W$ is also captured.
 Two minima in the conductance emerge for large enough junctions.
 In addition, our calculations yield reliable information on the doping variation along the junction for metallic
leads interacting weakly with graphene, and we find accurate potential profiles at the vicinity of the
metal boundary. We demonstrate that the dominant features of our first-principles results can be reproduced by an
analytically tractable effective model. The parameters of the effective model are derived from first-principles.
 As an application we use the effective model to predict the conductance in junctions with Au as the supporting metal at the leads.

We thank L. Xian, K. Park, and E. Yepez for helpful discussions. This work is supported by the Department of Energy (Grant No.
DE-FG02-97ER45632). We acknowledge interaction with the Georgia
Tech MRSEC funded by the National Science Foundation (Grant No.
DMR-02-05328) and computer support from Teragrid (TG-PHY090002) and NERSC.



\end{document}